# Evaluation of the effectiveness of an introductory mechanics Lab with Arduino and smartphone

Eugenio Tufino[1] and Giovanni Organtini[2]

[1]Department of Physics, University of Trento, Povo (Trento), Italy

[2]Sapienza Università di Roma & INFN-Sez. di Roma, Italy

**Abstract.** This article describes the reform of a physics undergraduate lab course by introducing new technologies such as Arduino microcontroller and smartphone, to help students experience a more authentic lab experience. This paper outlines the structure of the lab activities and presents findings from pre- and post-course questionnaires, including the E-CLASS survey, completed by participating students. Some interesting observations about teamwork and equity emerged from these questionnaires, particularly about female students. These results will guide us in designing more effective and equitable lab courses.

In this paper we describe the reform of a mechanics laboratory course with the introduction of digital data acquisition devices. The idea of this reform dates back a few years but some resistance had been encountered from the faculty, so its implementation was facilitated by the emergence of the COVID-19 pandemic that forced Universities to find alternatives to in-person labs. A useful review of the wide range of approaches proposed by physics laboratory instructors in laboratory courses during the pandemic is available here [1].
In general, in recent years, even before the pandemic, there has been a resurgence of interest in introductory physics lab courses also partly motivated by studies showing that students do not effectively learn physics concepts in traditional (confirmatory type), highly structured labs [2,3]. We revised the overall approach of the course to introduce some active learning strategies, giving more space for students to make decisions and reducing the verification aspects in lab activities. These in fact represent two of the most critical aspects of any lab course [4].
The use of Arduino microprocessors represents a pedagogical support, students in fact were encouraged to build their own apparatus from scratch and decide on strategies for making measurements. The changes were implemented gradually. The 2021 edition, due to the pandemic, was held mostly remotely with students doing experiments from home in an active and collaborative approach with breakout rooms. The course lectures were given by a single lecturer, while for the laboratory part the students were divided into two groups, one (pilot lab) in which Arduino and smartphones were used for data acquisition, and Python language with Google Colab to do data analysis, while the other group used more traditional methods for data acquisition and data analysis. In the 2022 edition, the lectures and the lab activities were both run by one of the authors and held in person in the laboratory. To assess the lab transformation, we have applied the E-CLASS survey [5]. The survey is intended to measure students' views on the nature of experimental physics, the role of experimentation in physics, and their confidence in their ability to conduct experiments. E-CLASS allows tracking of changes in students' attitudes and beliefs as a result of the course intervention, as well as to identify areas where students may need more support.

To probe student perception of the laboratory classroom experience, an additional survey was administered at the end of the course. We also investigated group work in relation to gender. In this article, we report the details of the structure of the lab course and show the results on students' attitudes. Further details about this work can be found in ref. [6].

## 1 Course Structure

The introductory laboratory course held at University Sapienza of Roma focuses on selected topics of Mechanics. Apart from the laboratory activity, there are also some lessons on measurements and uncertainties, data analysis, probability theory and Bayesian reasoning.

Seven experimental activities are planned during the semester course, two of which are conducted individually, while the others are carried out in groups. They are outlined in Table 1 and are common to all the channels of the course, both "traditional" ones (traditional in the sense that they used traditional and standard instruments for experiments) and the revised one. During the lab sessions in groups, students work in teams made of three or four students. At the end of the lab, they have to submit a lab report, which is reviewed with feedbacks and returned by the instructors before the next lab session.

Table 1. Experiments proposed in the 2021/2022 course with different technologies: Arduino (A), smartphone (S), dedicated devices (D)

| Experiments activities in 2021/22 | |
|---|---|
| Measuring the density of a body | D |
| Studying the motion of a pendulum | S |
| Studying the dynamics of a spring | A |
| Finding how the speed of sound depends on temperature | A |
| Radiation counters | D |
| Measuring the moment of inertia of a rolling cylinder | S |
| Studying the deformation of a slab | A |

### 1.1 Arduino and smartphone

Arduino is a popular open-source platform and can be used very effectively in physics laboratories as a tool for data acquisition and data analysis. Considering the affordable price, we asked students to purchase a simple kit consisting of an Arduino UNO board, a breadboard, a few connecting cables, an ultrasonic sensor, and a temperature sensor. This allows students to use Arduino at home for further experiments on their own. Since students have already taken a C programming course in the months prior to the lab course, they already know the syntax of Arduino programming and so a session of only two hours was sufficient to focus on specific statements for reading analog and digital pins, setting digital pins, etc. A program to read data from an ultrasonic sensor was used as an application example. In addition to Arduino, the smartphone is also used in the course to acquire data. Specifically, the Phyphox application [7] is used, which is free, open-source and transforms smartphone sensors into measurement instruments. It can collect data from the accelerometer, gyroscope, magnetometer, video camera, microphone, light sensor and barometer.

To give some examples of the proposed activities, in the experiment on the dynamics of a spring, students used the ultrasonic sensor connected to Arduino (figure 1, left) to determine the position of the end of a suspended spring and then, using this data, they estimated the period T of oscillation. They then plotted $T^2$ as a function of the suspended mass to determine the elastic constant with a linear fit. In studying the dynamic of pendulum, the smartphone's accelerometer was used to measure the period of the smartphone, which was used as the suspended mass. The gravitational acceleration was determined by fitting the distribution of $T^2$ values to the pendulum length.

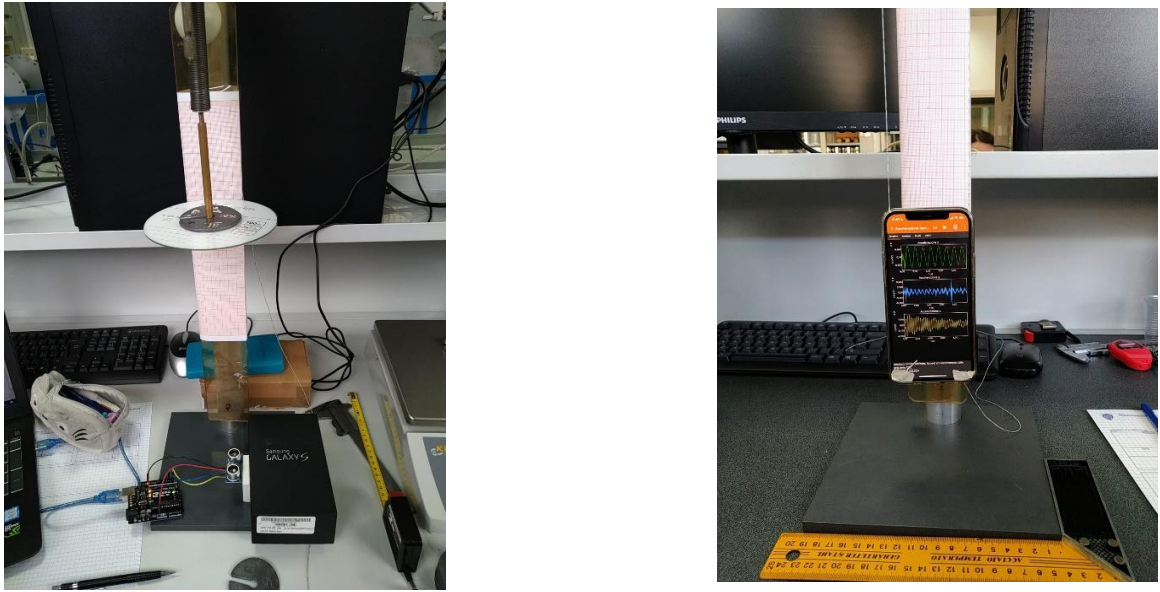

**Figure 1.** **(Left)** An example of an experimental apparatus with Arduino and ultrasound sensor to study the dynamics of a spring. **(Right)** Students' experimental setup for the pendulum experiment with smartphones and Phyphox.

Similarly, the study of the deformation of a slab can be done with either Arduino, measuring its deformation by means of the ultrasonic sensor (figure 2), or with a smartphone, fixed to the slab and oscillated. The period of oscillation is proportional to the cube of the length of the slab, as well as its deformation.

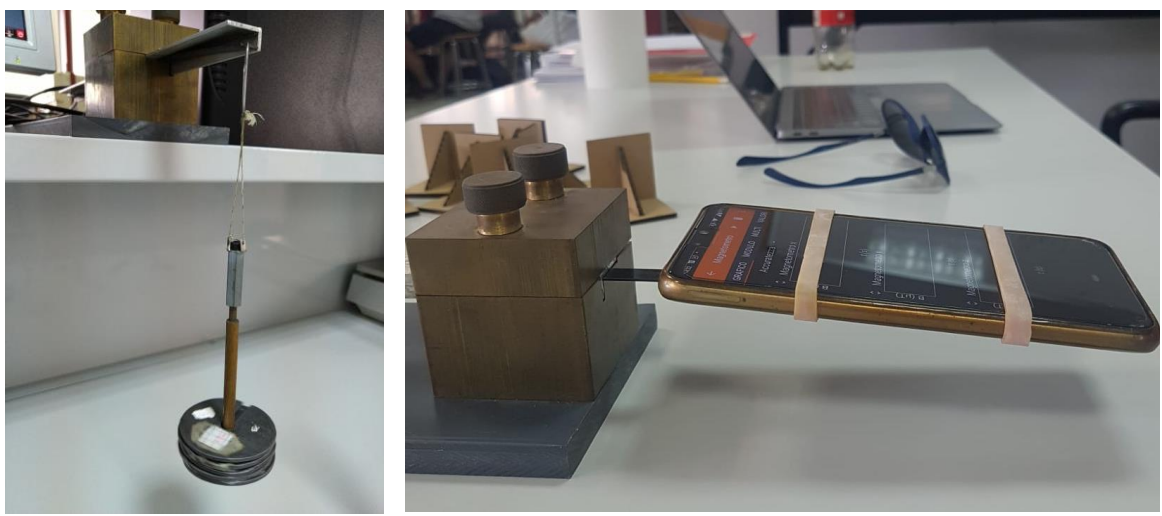

**Figure 2.** **(Left)** The deformation of a slab depends on the weight applied to its end. It is studied by measuring the modification of the height of the weights with an Arduino ultrasonic sensor. **(Right)** The period of oscillations of the slab depends on its length and is measured using a smartphone.

*1.2 Data Analysis with Jupyter Notebook*

For data analysis of experiments, students used Jupyter Notebook with Python language. Since the students had already taken a C programming course, the introduction to Python programming and Jupyter notebook use was very short in the first weeks of the course. Jupyter notebook allows users to write and execute code, create visualisations, insert text, images and videos, all in one place.

In particular, we proposed Google Colab [8], a cloud-based platform available to anyone with a Google Account. Thus, students do not have to do any installation-which can sometimes be problematic-and they can easily share the notebook in Google Drive and collaborate in group for data analysis and lab report writings.

Jupyter Notebooks were also a valuable resource during data analysis and probability classes, where the teacher used some notebooks to introduce complex concepts and simulation, as well as to provide hands-on examples and exercises for students. An example of this teaching method can be seen in the notebook available at the following link, which was used to propose an application of Bayes' theorem [9]. The use of Jupyter Notebooks addresses the recently emerged need to introduce computational skills as early as possible in the undergraduate courses. Examples of this include the AAPT Conference Report from 2019 [10] and the PICUP group [11].

*1.3 Pedagogical Approach*

The lectures typically revolved around explaining the principles of statistical data analysis, introduced first heuristically and then proceeding to a more formal justification of the practices used. This way we were promoting scaffolding. Students were introduced to Bayesian reasoning, fitting data with the least squares method, and the maximum likelihood principle. They learned how to estimate confidence intervals and p-values.

One of the intentions of the course is to help students understand that performing a measurement is always likened to drawing a random number with an appropriate probability distribution. We discussed the experiments and the way in which they are conducted only loosely, without too many details, to foster students' engagement in the preparation of them.

By utilising Arduino and smartphones instead of a dedicated standard experimental setup, students are encouraged to design the experimental apparatus, take decisions and in case of discrepancies between measurements and the mathematical model, work on improving the apparatus and clarifying the approximations or assumptions of the model.

The Arduino-based instrumentation is intended to be a tool for students to explore and understand. The objective is to move beyond treating the data acquisition system as a black box. Students are encouraged to experiment with editing both the data analysis code in the Jupyter notebook on the computer and the data acquisition code on the Arduino IDE. Before each lab session, students were sent a document containing general guidelines about each lab activity; these sheets were designed so as not to specify too many details, not following a step-by-step recipe format, leaving room for student creativity and initiative. Students are also given the autonomy to incorporate their own research question related to the experiment into the final report.

## 2 Methods and Data Collection

To evaluate the impact of the revised lab, we assessed students' attitudes using the E-CLASS survey, conducted observations, analysed student work and submitted additional questionnaires to students to understand how the proposal was received, including aspects related to group work.

*2.1 E-CLASS survey*

The E-CLASS survey is a research-based assessment tool composed of 30 Likert-like questions. Various aspects are probed, including confidence, experimental design, purpose of labs, statistical uncertainty, systematic error, troubleshooting. The E-CLASS assessment is described in this paper, along with a summary of the research conducted using it [12].

We translated into Italian (available on PhysPort website [13]) and administered it to students for the first time in the 2021 edition of the course both at the start and end of the semester. Recently, the E-CLASS survey has been translated and used in German laboratory courses [14].

The E-CLASS requires students to provide two responses to the same question: one from their own point of view and one from the point of view of a professional physicist. These responses are labelled in figure 3 as YOU (the student viewpoint) and EXPERT (the expert viewpoint) and are analysed separately. The answers provided by the students are then compared with those provided by the experimental physicists, who serve as a reference for the EXPERT [5].

To calculate the score, we applied a two-point scheme, by awarding students 1 point for each question where their response matches the expert-like response and zero point for any other response. This means that the average score for any given question represents the percentage of students who gave a favourable response. The overall E-CLASS score is determined by summing the scores on all 30 questions, which represents the average number of favourable responses from each student.

### 2.2 Student Perception of the Lab experience

In the last week of the course, and before the final exam, we administered another anonymous questionnaire, referred to as the end-of-course questionnaire. This instrument developed following general guidelines outlined for example in [15], and further designed by us, contained a set of questions aimed at obtaining feedback from students about the perceived usefulness of the lab experience and its practical organization. The questionnaire also included questions specifically related to the use of Arduino in the lab. Most of the questions in the questionnaire can be viewed in figure 4, figure 5, and figure 6. The number of students who responded to the end-of-course questionnaire was 46, which is about 50 percent of those enrolled in the section of the course lab.

## 3 Student Learning: results

In this section we illustrate the results obtained with the tools described in Section 2.

### 3.1 Analysis of E-CLASS survey

In figure 3 we show the overall class score for the 2021/22 edition. For the YOU (student's point of view) questions, only a slight increase in the overall score, not statistically significant, was observed between the pre- and post-questionnaires. As you can see, the score for the student viewpoint towards EXPERT increased from 0.82 to 0.85. The latter result may indicate that through the laboratory course, knowledge of how physicists work has slightly improved in students.

The fact that there is no deterioration in the average E-CLASS score is a good indication of the methodology adopted in the course with the introduction of digital devices into the laboratory. In overly structured and guided lab courses normally a deterioration of 1.9 point in average is observed [2]. Other types of laboratory courses where there are no negative shifts in the average E-CLASS score are ISLE or SCALE-UP type lab courses [16]

The E-CLASS survey was administered in presence to students in the lab during the first week and the final week of the semester. The number of participants was 88, and students chose a nickname to maintain their anonymity.

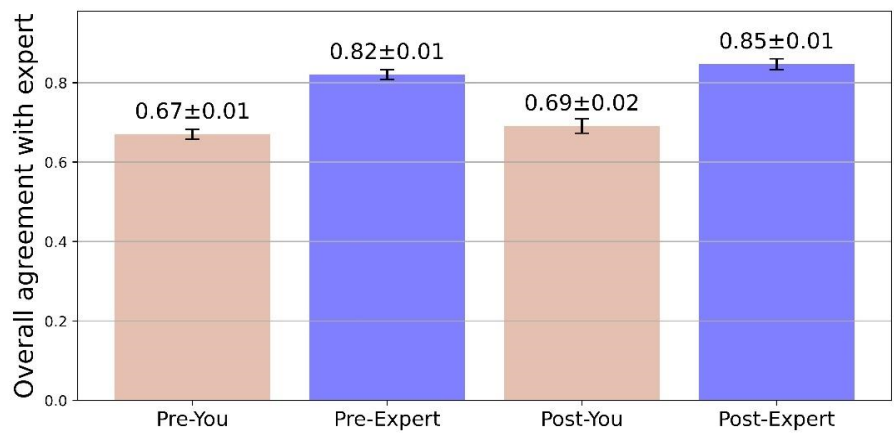

**Figure 3.** Average fractions of questions with favourable responses ("You" and "Expert" views) both before and after lab activities for year 2021/22.

More than for shifts in mean values, the most interesting indications provided to instructors by the questionnaire concern shifts (positive or negative) in individual questions. The questions for which we measured the statistically significant shifts obtained using the Mann-Whitney U-test, are:
- Q1 (When doing an experiment, I try to understand how the experimental setup works), from 0.77 to 0.94.
- Q8 (When doing an experiment, I try to understand the relevant equations) from about 0.73 to about 0.94.
- Q10 (Whenever I use a new measurement tool, I try to understand its performance
- limitations) from 0.53 to 0.74.

These three results seem to follow from the intentional and continued use of Arduino and smartphone to carry out the experiments. In addition, some questions kept their initial low scores, which means those aspects are not addressed sufficiently during the course to induce change. These include: Q4 (If I am communicating results from an experiment, my main goal is to have the correct sections and formatting), Q17 (When I encounter difficulties in the lab, my first step is to ask an expert, like the instructor) and Q29 (If I don't have clear directions for analysing data, I am not sure how to choose an appropriate analysis method).

The result of item Q4 seems to indicate that students misunderstand the demands on the lab report, so in future editions we will try to discuss with them more explicitly the characteristics of a good lab report and what to pay attention to. The results of Q17 and Q19 seem to indicate that low confidence and lack of autonomy in decision-making persist. Changing these convictions likely takes more than just the duration of one course.

### 3.2 *Student Perception of the Lab experience*
Referring to the end-of-course questionnaire described in the previous section, our focus was on understanding student perceptions of the reformed lab course. Figure 4 illustrates the responses to six of the questions from this questionnaire that relate to the student's perception and the use of Arduino.

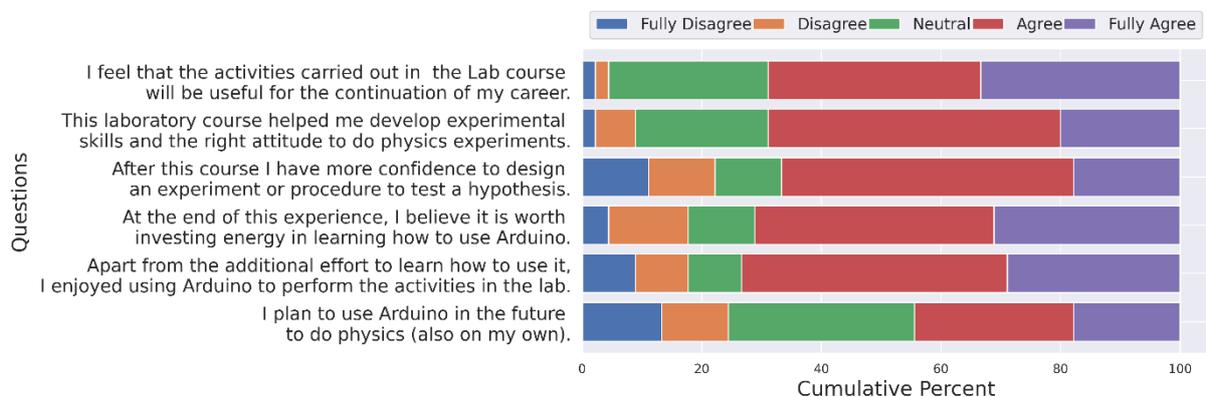

**Figure 4.** Stacked bar plots showing student responses to survey administered at the end of the course.

The analysis of student feedback revealed that most students have a favourable perception of their laboratory experience and feel that it has been important for the development of their skills. Students generally responded positively to the introduction of Arduino boards; some of them expressed excitement about the potential for the device and expressed interest in continuing to use Arduino in the future also on their own.

### 3.2.1 Roles in Lab activities and gender equity

In the two editions of the course, spontaneous formation of group members was not allowed, primarily to accustom students to working with people they may not know, as is common in professional settings. In the same survey we asked about the preferred roles during group lab work. While not mandatory, we recommended that each group rotate roles for each experimental session to give each group member the chance to take on different roles. Therefore, we also asked in the end-of-course questionnaire which role distribution model was preferred. Although the questionnaire was anonymous, we requested that participants indicate their gender. Of the 46 students who responded, 28 were male, 17 female, and one did not respond to the question.

Analysing the responses according to gender showed some interesting aspects. Contrary to popular belief, we found that gender is not a significant factor in the selection of roles for those responsible for building the equipment and collecting data. Males show a slight preference for programming compared to females, while, surprisingly, they seem to prefer the role of data analysis and selection of appropriate data representation more than females do, contradicting stereotypes (figure 5).

In general, about the role distribution scheme, females prefer a collaborative approach where everyone works together, while males tend to prefer a rotating system where roles are assigned among group members, as shown in figure 6. The findings align with previous research [17] and provide a valuable foundation for further studies in the field of physics education. Other interesting insights came from the comments written by students (see the following section).

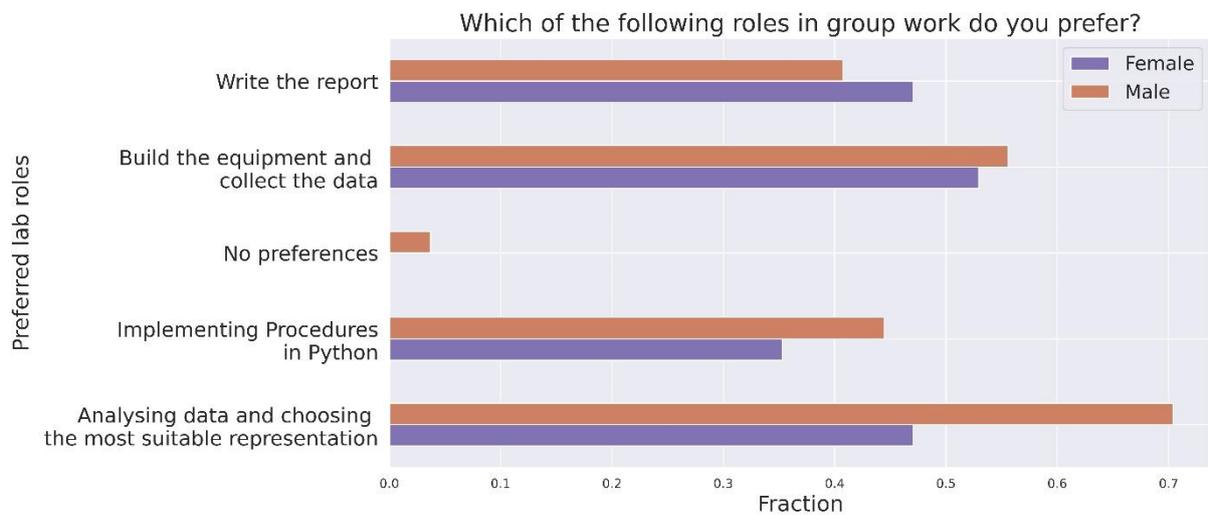

**Figure 5.** Student responses to survey items about preferred lab roles

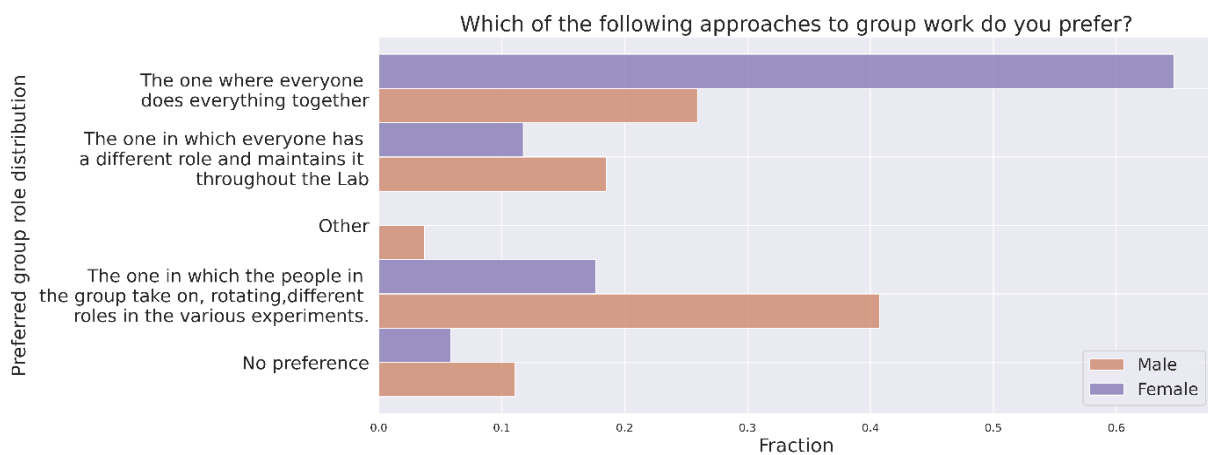

**Figure 6.** Student responses to survey items about preferred lab group distribution of work.

*3.2.2 Student Perspectives*

The last question in the end-of-course questionnaire asked students to make explicit the aspects of the lectures and laboratory activities that they perceived as problematic, both from the point of view of content and the teaching approaches. Some students highlighted the problem of time management, especially the time allowed for the delivery of lab reports. In the first lab sessions, students were asked to submit the report at the end of the corresponding lab session. Following feedback from students, the teacher adjusted this requirement, providing a 24-hour submission window.

According to some students' feedback, accommodating their request for more time was greatly appreciated by them and significantly improved their attitude towards the course. This underlines the importance for a teacher to listen to (and ponder) students' requests and show interest in their difficulties. Some students' comments refer to the composition of the group, and remarkably, all these comments come from female students, for example:

> *"As for the lab activities, by not allowing freedom of choice of groups, I have found in my group and others a great imbalance in the knowledge of their respective members. As for the four hours to accomplish the activity, I don't disagree, only that for groups with only one or two individuals who meet the required knowledge base, it becomes hell (I am referring to the unbalanced groups)" (translated)*

> *"My group is very uneven, I'm the only one doing anything, and I don't think staying up all night trying to come up with interesting conclusions would be a plausible option now or in the future. Group exercises only created stress and a desire for me to quit, because I never felt stimulated by my group mates and, rather, just left to myself (literally)." (translated)*

> *"The assortment of the group made it impossible to divide roles equally, leading one person to do the work of four people" (translated)*

Such imbalances in groups, like being left alone to do all the work, seems to have had a more detrimental effect on female students. Perhaps this is due to a lower level of confidence in female students.
To promote fairer and more equal group work in future editions we will allow students the freedom to form groups while requiring gender balance. We'll also make an effort to frequently ask students how their group work is progressing. Regarding the teaching methods in theory lessons on probability and data analysis, some students appreciated the approach designed to promote students learning on their own. Two female students pointed out the following:

> *"However, I think the most important lesson I learned from this course was to "jump in": to formulate hypotheses and try to test them rigorously, in a sense to believe in myself and my abilities (even if we were often wrong)." (translated)*

> *"…Nothing to report except in the fact that bringing all the lessons together in order to construct a proper logical thread was not a simple matter…" (translated)*

A male student reported:

> *"…I really appreciated the professor's approach to teaching the subject, applying and drawing on Physics education sources to propose more effective teaching methods. I also appreciated the beginning of the lectures helpful in contextualizing the purpose of studying physics and what it is for…" (translated)*

In contrast, three other students expressed a desire for more formal and systematic lectures:

> *"There is no clear support material to go along with the classroom lectures" or "For the future, I think more step-by-step guidance would be beneficial..." (translated)*

> *"During the study of probability, not carrying out problems in class in a progressively related way with the theory disadvantages its consolidation" (translated)*

> *"As for lectures, however, I think they need to be supplemented with PowerPoints or multimedia presentations of different kinds" (translated)*

The latest comments seem to indicate that students are accustomed to the traditional lecturing mode, where the teacher explains very systematically, and they take notes.
As reported in study [18] on measuring actual learning vs. feeling of learning in active learning and passive lecture classrooms, students who are exposed to active learning methods may underrate their own knowledge, but actually perform better than their peers who were taught using traditional lecturing methods. In a future edition of the course, when this request for lecturing emerges from some students, we plan to propose a discussion of this article, to encourage a change of expectation.

### 3.3    Distribution of students' final exam scores

Analysis of overall student scores on the final exam of the 2020/2021 and 2021/2022 editions [6] reveals no measurable difference between students in the Arduino/smartphone lab section and students enrolled in the "traditional" lab sections. This result indicates that although students had to learn how to use Arduino and smartphone and had to spend time making decisions on how to carry out the various proposed experimental activities, the reformed lab format did not put students at a disadvantage in the course.

During the final assessment of the 2021/22 academic year, although the distribution of scores was not greatly different from traditional sections, one of the authors, the teacher of the course, noted that students initially struggled with the course demands.

However, as the course progressed, they began to fully grasp the tools of their lab work and displayed a deeper understanding than in previous years, resulting in some cases, in excellent outcomes. Students evaluated the limitations and elaborated more critically and extensively on the answers. This aligns with findings on active learning - students need time to adjust to more open-ended teaching methods, and this can be facilitated by providing the appropriate support and guidance.

## 4    Conclusions

In this paper, we have described the transformation of a first-year physics laboratory course, which was achieved by introducing digital acquisition devices such as Arduino and the smartphone and elements of active learning in both theory lessons and laboratory work. This transformation gave students the freedom to design the experimental apparatus and analyse the data.

The results obtained from the E-CLASS responses of students show that this transformation is possible without a worsening of students' beliefs and expectations; E-CLASS allowed us to identify aspects in which there was improvement from the beginning of the course and areas in which students' perspectives remained far from those of the experts. The end-of-course questionnaire on students' experience indicated that students enjoyed working with Arduino.

The responses to this questionnaire revealed some interesting aspects about teamwork and equity with respect to gender, that is an interesting research topic. For example, it shows that women much more want to work together with group members by sharing all tasks. A subset of students would also benefit from additional support and scaffolding.

Considering the examination results, we obtained encouraging outcomes. Our conclusion is that active learning strategies and hand-on activities with Arduino and smartphone are beneficial in enhancing students' soft skills, fostering creativity and critical capabilities. In subsequent editions of the course we will try to take these implications into account.

## 5    Acknowledgments

Giovanni Organtini thanks Guido Fantini and Adriano Frattale Mascioli for their commitment running the laboratory classes. Eugenio Tufino acknowledges Micol Alemani and Michael F.J. Fox for useful discussions and thanks Stefano Oss for his support. We are indebted to retired Franco Meddi, who took this course till 2020/21 who gave us the possibility to test the test the methods and for his invaluable insights.